\documentclass[pre,twocolumn,showpacs,preprintnumbers,amsmath,amssymb]{revtex4}      
\usepackage{epsfig}%
\usepackage{longtable}
\usepackage{eucal}
\begin{document}
\setlength{\topmargin}{-0.25in}
\preprint{U. of IOWA Preprint}

\title{Lattice gluodynamics at negative $g^2$}


\author{L. Li and Y. Meurice}
\email[]{yannick-meurice@uiowa.edu}
\affiliation{Department of Physics and Astronomy\\ The University of Iowa\\
Iowa City, Iowa 52242 \\ USA
}


\date{\today}

\begin{abstract}

We consider Wilson's $SU(N)$ lattice gauge theory (without fermions) at negative 
values of $\beta= 2N/g^2$ and for $N$=2 or 3. We show that in the limit $\beta 
\rightarrow -\infty$, the path integral is dominated by configurations where links variables are set to 
a nontrivial element of the center on selected non intersecting lines. For $N=2$, these configurations can be characterized by a unique gauge invariant set of variables, while for $N=3$ a multiplicity growing with 
the volume as the number of configurations of an Ising model is observed. In general, there is a discontinuity in the 
average plaquette when $g^2$ changes its sign which prevents us from having a convergent series in $g^2$ 
for this quantity.
For $N=2$, a change of variables relates the gauge invariant observables at positive and negative values of $\beta$. 
For $N=3$, we derive an identity relating the observables at $\beta$ with those at $\beta$ rotated by $\pm 2\pi/3$ in the complex plane and show numerical evidence for a Ising like first order phase transition near $\beta=-22$. We discuss the possibility of having lines of first order phase transitions ending at a second order phase transition in an extended bare parameter space.

\end{abstract}


\maketitle

\section{Introduction}
It has been known from the early days of QED that perturbative series have a zero 
radius of convergence \cite{dyson52}. This has not prevented Feynman diagrams to become 
an essential tool in particle physics. However, perturbative series need to be used 
with caution. The divergent nature of QED series was foreseen by Dyson as a consequence of the 
apparently pathological nature \cite{dyson52} of the ground state in a fictitious world with negative $e^2$. Like charges then attract and pair creation can be invoked to produce states where electrons are brought together in a given region and positrons
in another. Dyson concludes that as this process sees no end, no stable vacuum can exist.

For Euclidean lattice models, related situations are encountered.
For scalar field theory with $\lambda \phi^4$ interactions, configurations with large 
field values make the path integral ill-defined when $\lambda < 0$ (provided that no higher 
even powers of $\phi$ 
appear in the action with a positive sign and that the path of integration is not modified). 
Modified series with a finite radius of convergence can be obtained 
by introducing a large field cutoff \cite{pernice98,convpert}. We are 
then considering a slightly different problem. In simple situations \cite{optim}, it is possible to determine an optimal value of the field cutoff that, at a given order 
in perturbation, minimizes or eliminates the discrepancy. 
For non-abelian gauge theories in the continuum Hamiltonian 
formulation, the substitution $g\rightarrow ig$ makes the quartic part unbounded from below and the 
cubic part non-hermitian. 

It should be noted that in quantum mechanics\cite{bender}, it is possible to change the boundary conditions of the
Schr\"{o}dinger equation in such a way that the spectrum of an harmonic oscillator 
with a perturbation of the 
form $ix^3$ or $-x^4$ stays real and positive.  The procedure can be extended to scalar field theory  in order to define a sensible $i\phi^3$ theory \cite{bender04}. 
Even though
conventional Monte Carlo calculations would fail for these models, complex Langevin methods can be used to
calculate Green's functions \cite{bernard2001}.

In the case of lattice gauge theory with {\it compact} gauge groups, the 
action per unit of volume is bounded from below and there is no large 
field problem. Consequently, these models have well-defined expectation 
values when $g^2<0$ and we can consider the limit $g^2\rightarrow 0^-$.
In this article, we discuss the behavior of Wilson loops for $SU(N)$ lattice gauge theory with $g^2<0$.
This work is motivated by the need to understand the unexpected behavior of the lattice pertubative series 
for the $1\times 1$ plaquette calculated up to order 10 \cite{alles93,direnzo95,direnzo2000}. 
An analysis of the successive ratios \cite{rakow2002,lilipro} may suggest that the series has a finite 
radius of convergence and a non-analytic behavior near $\beta\simeq 5.7$ in contradiction with 
the general expectations that the series should be asymptotic and the transition from weak to strong 
coupling smooth.

We consider here pure (no fermions) 
gauge models with a minimal lattice action \cite{wilson74c}. For definiteness this model and our notations are defined in section \ref{sec:model}. 
The extrema of the action are discussed in \ref{sec:limit} and enumerated for $SU(2)$ and $SU(3)$.
We then discuss (section \ref{sec:su2}) 
the case of 
$SU(2)$ and show that planar Wilson loops with an area $A$ (in plaquette units) pick up a factor $(-1)^A$ when 
$g^2$ becomes negative and the behavior for $g^2<0$ is completely determined by the behavior with $g^2>0$.
As the Wilson loops are nonzero when $g^2\rightarrow 0^+$, the ones with an odd area have a discontinuity 
which invalidates the idea of a regular perturbative series. 

The case of $SU(3)$ is discussed in section 
\ref{sec:su3} where we derive identities involving Wilson loops 
calculated with a coupling rotated by $\pm 2\pi/3$ in the complex 
plane. Consequently, for $N=3$, we have no a-priori knowledge regarding the behavior of Wilson loops 
when $g^2<0$. 
We report numerical evidence for a first order phase transition near $\beta=6/g^2\simeq -22$ using 
methods similar to Ref. \cite{creutz81}. The implications of our findings are summarized in the conclusions.

\section{The model, notations}
\label{sec:model}
In the following, we  consider the minimal, unimproved, lattice gauge model originally 
proposed by K. Wilson \cite{wilson74c}. 
Our conventions and notations are introduced in this section
for definiteness. 
We consider a cubic lattice in $D$ dimensions.  A $SU(N)$ group element is attached to each link $l$ and $U_l$ denotes its fundamental representation. $U_p$ denotes the conventional product of $U_l$ (or  hermitian conjugate) along the sides of a 
$1\times 1$ plaquette $p$.
The minimal action reads
\begin{equation}
S=\beta\sum_{p}(1-(1/N)Re Tr(U_p)) \ ,
\end{equation} 
with $\beta=2N/g^2$.
The lattice functional integral or partition function is 
\begin{equation}
Z=\prod_{l}\int dU_l {\rm e}^{-S} \,
\end{equation}
with $dU_l$ the $SU(N)$ invariant Haar measure for the group element associated with the link $l$. 
The average value of any quantity $\mathfrak{Q}$ is defined as usual by inserting 
$\mathfrak{Q}$ in the integral and dividing by $Z$.

In the following, we consider symmetric finite lattice with $L^D$ sites and periodic boundary conditions. 
For reasons that will become clear in the next sections, $L$ will always be even.
The total number of $1\times 1$ plaquettes is denoted 
\begin{equation}
\mathcal{N}_p\equiv\ L^D D(D-1)/2\ .
\end{equation}
Using
\begin{equation}
f\equiv-(1/\mathcal{N}_p)\ln Z\ ,
\end{equation}
we define the average density
\begin{eqnarray}
\label{eq:pdef}
P(\beta)&\equiv & \partial f/\partial \beta \nonumber \\ 
&=&(1/\mathcal{N}_p)\left\langle \sum_p
(1-(1/N)Re Tr(U_p))\right\rangle \ .
\end{eqnarray}
In statistical mechanics, $f$ would be the free energy density multiplied by $\beta$ and 
$P$ the energy density. In analogy we can also define the constant volume specific heat per plaquette
\begin{equation}
\label{eq:cv}
C_V=-\beta^2 \partial P/\partial \beta \ .
\end{equation}

\section{The limit $\beta \rightarrow -\infty$}
\label{sec:limit}
In the limit $\beta \rightarrow -\infty$, we expect the functional integral to be dominated 
by configurations which {\it maximize} $ \sum_{P}(1-(1/N)Re Tr(U_P))$. 
In the opposite limit ($\beta \rightarrow +\infty$), the same quantity needs to  
be minimized which can be accomplished by taking $U_l$ as the identity everywhere.

We first consider the question of finding the extrema of $TrU$. For our study of the behavior when 
$\beta \rightarrow -\infty$, we are particularly interested in finding absolute minima of $TrU$.
Using $TrU=Tr(VUV^{\dagger})$ for $V$ unitary, $U={\rm e}^{iH}$ with $H$ traceless and hermitian, and $V{\rm e}^{iH}V^{\dagger}$=${\rm e}^{iVHV^{\dagger}}$, we can diagonalize $H$ and write 
\begin{equation}
ReTrU=\sum_{i=1}^{N-1}\cos(\phi_i)+\cos(\sum_{i=1}^{N-1}\phi_i)\ .
\end{equation}
The extremum condition then reads
\begin{equation}
\label{eq:extr}
\sin(\phi_i)+\sin(\sum_{i=1}^{N-1}\phi_i)=0\ ,
\end{equation} 
for $i=1,\dots N-1$.
The trivial solution is all $\phi_i=0$. We then have $ReTrU=N$ which is an absolute maximum. 

For $N=2$, we have only one nontrivial solution $\phi_1=\pi$, which corresponds to the nontrivial element of the center $U=-\openone$. 
We then have $ReTrU=-2$ which is an absolute minimum. 

For $N=3$, we have 5 nontrivial solutions. Two correspond to the nontrivial elements of the center ($\phi_1=\phi_2=\pm 2\pi/3$). The matrix of second derivatives has two 
positive eigenvalues and these two solutions correspond to a minimum. 
We use the notation $\Omega\equiv {\rm e}^{i2\pi/3}\openone$ on the diagonal. We have $ReTr\Omega=-3/2$, 
which we will see is an absolute minimum. The other three solutions are $(\phi_1=\pi,\phi_2=0)$, $(\phi_1=0,\phi_2=\pi)$ and $(\phi_1=\pi,\phi_2=\pi)$. They 
correspond to elements conjugated to diagonal matrices belonging to the three canonical $SU(2)$ subgroups with the $SU(2)$ element being the non trivial center element. These three solutions have matrices of second derivatives with eigenvalues of
opposite signs and correspond to saddle points rather than minimum or maximum. In the 
three cases $ReTrU=-1$. 

For general $N$, it is clear that we can always find at least one group element $U$ 
such that $ReTrU$ is an absolute minimum. In particular, for $N$ even, $U=-1$ is such 
a group element, with $ReTrU=-N$ (the individual matrix elements must have a complex norm less then one). 
For $N\geq 3$ and odd, it is easy to check that all $\phi_i=(N-1)\pi/N$ is a solution of the extremum condition Eq. (\ref{eq:extr}). For this choice, $ReTrU=-N|\cos((N-1)\pi/N)|$ which is clearly negative and tends to $-N$ as 
$N$ becomes large. This solution (the element of the center the closest to $-\openone$) gives an absolute 
minimum of $TrU$ for $N=3$ and we conjecture that it is also the case for larger $N$.

We can now obtain an absolute minimum of the action if we can build a configuration such that $ReTrU_P$ takes its absolute minimum value for every plaquette.  This can be 
accomplished by the following construction.
In the Appendix, we show that (at least for for $D\leq 4$) and for $L$ even, it is possible to construct a set of 
lines on the lattice such that every plaquette shares {\it one and only one} link with this set of lines. 
We call such a set of links $\mathfrak{L}$.
One can then put an element which gives an absolute minimum of $ReTrU$ 
on the links of $\mathfrak{L}$ and the identity on all the other links. For $SU(2)$, there is only one possible choice that minimizes $ReTrU$, namely $U=-\openone$. For $SU(3)$, there are two possible choices 
$U=\Omega$ or $U=\Omega^{\dagger}$. We emphasize that the construction only works for $L$ even. If $L$ is odd, 
there will be lines of frustration in every plane.

The set of links $\mathfrak{L}$ is not unique. Starting with a given set, we can generate another one 
by translating the lines by one lattice spacing or rotating them by $\pi/2$ about the lattice 
axes. By direct inspection, it is easy to show that for $D=2$ there are 4 such a sets of lines while 
for $D=3$ there are 8 of them. 

Enumerating all the gauge inequivalent minima of the action at negative $\beta$ for arbitrary $D$ and 
$N$ appears as a nontrivial problem. 
In the rest of this section, we specialize the discussion to $N=$ 2 or 3. In order to discriminate 
among  
gauge inequivalent configurations, it
is useful to make the following (gauge invariant) argument: 
in order to have an absolute minimum of the action, for every 
$1\times1$ plaquette $p$, the product $U_p(\mathbf{n})$ of the $U_l$ along $p$ starting at any site 
$\mathbf{n}$ of $p$, is a nontrivial element of the center. Under a local gauge transformation, 
$U_p(\mathbf{n})\rightarrow V(\mathbf{n})U_p(\mathbf{n}) V(\mathbf{n})^{\dagger}=U_p(\mathbf{n})$ since $U_p(\mathbf{n})$ commutes with any $SU(N)$ matrix. For the same reason, changing ${\mathbf{n}}$ along the plaquette amounts to a $VUV^{\dagger}$ conjugation and has no effect on the center. 
Consequently, configurations with a different set of $\mathfrak{U}=\{ U_p \}$ are not gauge equivalent. 
One can think of $\mathfrak{U}$ as a set of electric and magnetic field configurations.

For $SU(2)$, there is only one, uniform, set 
$\mathfrak{U}$ where all the elements $U_p=-\openone$. For $D=2$, this can be realized in 4 different ways 
by putting $-\openone$ 
on the 4 distinct sets $\mathfrak{L}$. These 4 configurations are all gauge 
equivalent. The gauge transformations that map these 4 configurations into each others can be obtained 
by taking $V=-\openone$ on every other sites of the lines of $\mathfrak{L}$.
For $D=3$, it is also possible to show that the 8 configurations that can be constructed with a similar 
procedure can also be shown to be gauge equivalent. The gauge transformation 
can be obtained 
by taking $V=-\openone$ on every other sites of the lines of $\mathfrak{L}$ pointing in two particular directions 
and in such way that one half of the lines created by the gauge transformation associated with one direction 
``annihilate'' with one half of the lines created by the gauge transformation associated with the other direction. 
We conjecture that in higher dimensions, the configurations that 
minimize the action for $SU(2)$ are also related by gauge transformations.

For $SU(3)$ the situation is quite different because for every link of a particular $\mathfrak{L}$, we have two possible 
nontrivial element of the center. Since there are $\mathcal{N}_p=L^D D(D-1)/2$ plaquettes on a $L^D$ lattice and one link of $\mathfrak{L}$ per plaquette, shared by $2(D-1)$ plaquette, we have $DL^D/4$ links in any $\mathfrak{L}$. 
Picking a particular $\mathfrak{L}$, it possible to construct $2^{DL^D/4}$ distinct $\mathfrak{U}$. Consequently, 
there are at least $2^{DL^D/4}$ gauge inequivalent minima of the action for $SU(3)$. Note that  $2^{DL^D/4}$ always 
is an integer for $L$ even, which has been assumed. In the case $D=4$, the degeneracy is simply $2^{L^4}$ which is the 
same as the number of configurations of an Ising model on a $L^4$ lattice.

In summary, we predict a discontinuity in $P$ as $g^2$ changes sign. In the limit $\beta \rightarrow +\infty$, 
we have $P\rightarrow 0$, while in the limit $\beta \rightarrow -\infty$, we expect $P\rightarrow 2$ for $N$ even,  
and $1+|\cos((N-1)\pi/N)|$ for $N$ odd.

\section{\label{sec:su2}N=2}

\begin{figure}
\includegraphics[width=2.5in,angle=0]{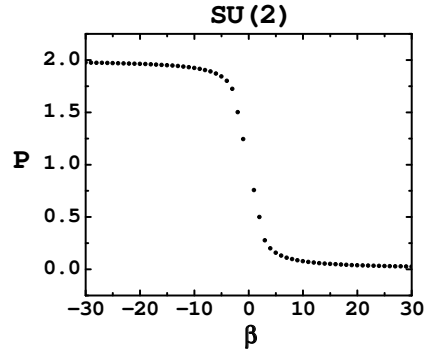}
\caption{\label{fig:su2pp}The average action density $P(\beta)$ for $SU(2)$.}
\end{figure}

\begin{figure}[ht]
\includegraphics[width=2.8in,angle=0]{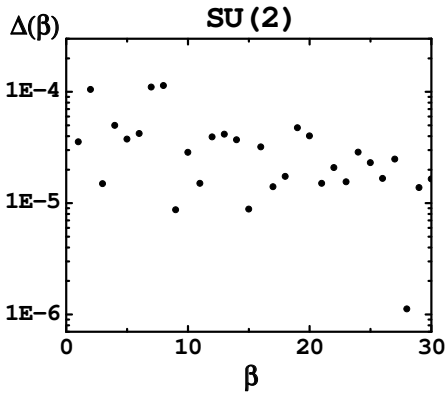}
\caption{\label{fig:pdelta}$\Delta(\beta)$ defined in Eq. (\ref{eq:delta}) versus $\beta.$}
\end{figure}

In this section we discuss $SU(2)$ gauge theories at negative $\beta$. The basic idea is that it is possible to change $\beta Re TrU_p$ into $-\beta Re TrU_p$ by making the {\it change of variables} 
$U_l\rightarrow -U_l$ for every link $l$ of a particular $\mathfrak{L}$. Since $-\openone$ is an element 
of $SU(2)$ and since the Haar measure is invariant under left or right multiplication by a 
group element, this does not affect the measure of integration. 
Consequently, we have
\begin{equation}
	Z(-\beta)={\rm e}^{2\beta\mathcal{N}_p}Z(\beta)
	\label{eq:su2id}
\end{equation}
Taking the logarithmic derivative as in Eq. (\ref{eq:pdef}), we obtain
\begin{equation}
\label{eq:pp2}
	P(\beta)+P(-\beta)=2\ .
\end{equation}
This identity can be seen in the symmetry of the curve $P(\beta)$ shown in Fig. \ref{fig:su2pp}. 
The validity of Eq. (\ref{eq:pp2}) can be further checked by calculating the difference 
\begin{equation}
\label{eq:delta}
\Delta(\beta)\equiv |	 P(\beta)+P(-\beta)-2|\, 
\end{equation}
which should be zero except for statistical fluctuations. 
Fig. \ref{fig:pdelta} illustrates this statement and shows that the 
statistical errors of our calculations are of order $10^{-4}$ or less.

The relation between $P(\pm\beta)$ of Eq. (\ref{eq:pp2}) together with the assumption that $P(+\infty)=0$, 
is in agreement with the statement made in section \ref{sec:limit} that $P$ seen as a function of $g^2=2N/\beta$, jumps discontinuously by 2 as 
$g^2$ becomes negative. This invalidates the idea that $P$ could have a regular expansion about $g^2=0$ with a non-zero radius of convergence. 

This relation can also be used in the opposite limit and expanded about $\beta =0$.
The odd terms cancel automatically. The even terms of order 2 and higher add and cannot cancel. Consequently, the even coefficients of the strong coupling expansion of $P(\beta)$ and the odd coefficients of the free energy should vanish, in agreement with explicit calculations \cite{balian74}.

The discontinuity at $g^2=0$ can be extended to Wilson loops of odd area (in plaquette units).
To see this, let us consider a Wilson loop $W(C)$ with $C$ a contour that is the boundary of an area 
made out of 
$A$ plaquettes. For simplicity, let also assume that this area is connected and has no
self-intersections.  Under the change of variables  $U_l\rightarrow -U_l$ for every link $l$ of an arbitrary set $\mathfrak{L}$, we have $W(C)\rightarrow (-1)^{A}W(C)$. This follows 
from the fact that for any line, 
the parity of the number of links of $C$ shared with this line, is the same as the number of plaquettes 
of the area in contact with this lines. Since $\mathfrak{L}$ shares a link with 
every plaquette, we obtain the desired 
result. This can be summarized as 
\begin{equation}
\left\langle W(C) \right\rangle	_{-\beta}=(-1)^{A}\left\langle W(C) \right\rangle	_{\beta}
\label{eq:parity}
\end{equation}

We can now try to interpret the change of the Wilson loop with the area in a term of a potential.
We consider a rectangular $R\times T$ contour $C$ and write
\begin{equation}
W(R,T,\beta)\equiv\left\langle W(C) \right\rangle	_{\beta}\propto {\rm e}^{-E(R,\beta)T}\ .
\end{equation}
From Eq. (\ref{eq:parity}) this implies
\begin{equation}
E(R,-|\beta|)	=E(R,|\beta|)+i\pi R \ .
\end{equation}
This property can be related to the fact that the configurations of minimum action are invariant under translations
by two lattice spacings but not  under translations
by one lattice spacing. This also confirms our expectation that the hamiltonian develops a nonhermitian part.
\section{N=3}
\label{sec:su3}
For $N=3$, $-\openone$ is {\it not} a group element and the closest thing to the change of variables used for 
$N=2$ that we can invent is a multiplication by a nontrivial element of the center $\zeta \openone$ for the links 
of a particular set $\mathfrak{L}$. We then obtain 
\begin{eqnarray}
Z(\zeta\beta)&=&{\rm e}^{(1-\zeta)\beta\mathcal{N}_p}Z(\beta)\nonumber \\ 
&\times& \left\langle {\rm e}^{(\beta/3)\sum_p(\zeta Re \zeta^{\star}Tr U_p-ReTrU_p)}\right\rangle _{\beta}  \ .
\end{eqnarray}
In the case $N=2$, $\zeta $ is replaced by -1, $\left\langle ...\right\rangle _{\beta}$ becomes 1 and 
we recover Eq. (\ref{eq:su2id}). In the case of $SU(3)$, 
the factor $\left\langle ...\right\rangle _{\beta}$ prevents us from deriving an 
exact identity analog to 
Eq. (\ref{eq:pp2}) for $SU(2)$. It is however possible to obtain an approximate generalization which is a 
good approximation for small $\beta$.
Setting $\beta=\zeta^{\star}x$, 
taking the logarithmic derivative with respect to $x$ and setting $x=\zeta \beta$, we obtain
\begin{equation}
P(\zeta \beta)=1-\zeta^{\star}+\zeta^{\star}P(\beta) + {\mathcal O}(\beta) \ .
\end{equation}
Taking the real part and using $1+\zeta + \zeta^2 =0$, we obtain
\begin{equation}
\label{eq:ppp}
	P(\beta)+P(\zeta\beta)+P(\zeta^2\beta)=3+{\mathcal O}(\beta^6)\ .
\end{equation}
which can be seen as an approximate $SU(3)$ version of Eq. (\ref{eq:pp2}).
The cancellation of the terms of order 1, 2, 4 and 5 occurs independently of the values of the 
coefficients at these orders. The absence of contribution of order 3 and the presence of a nonzero 
contribution at order 6 comes from the fact \cite{balian74err} that $ln(Z)/{\mathfrak N}_p$ has a zero (nonzero) contribution at order 4 (7).

\begin{figure}[ht]
\includegraphics[width=2.8in,angle=0]{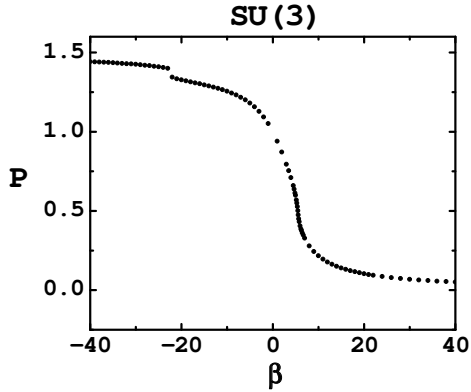}
\caption{The average action density $P(\beta)$ for $SU(3)$}
\label{fig:su3p}
\end{figure}

As it does not seem possible to obtain $P(\beta)$ for $\beta$ real and negative from our knowledge at $\beta$ real and positive, we have to resort to a direct numerical approach.
The results are shown in Fig. \ref{fig:su3p}. A discontinuity near $\beta=-22$ is clearly visible. This indicates a first order phase transition.

The metastable branches have been studied following the approach 
of M. Creutz \cite{creutz81} used to study of a fist order transition for $SU(5)$. 
As $\beta$ becomes more and more negative, the system becomes more ordered but has also 
higher average energy $P$, the supercooled/heated terminology may be confusing and will be avoided.

\begin{figure}[ht]
\includegraphics[width=2.8in,angle=0]{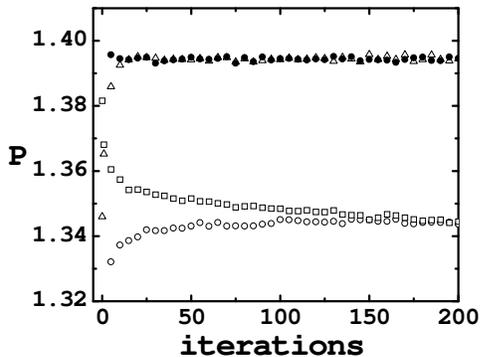}
\caption{$P$ as a function of iterations for the four initial configurations 
described in the text.} \label{fig:meta}
\end{figure}

We have run Monte Carlo simulations on a $8^4$ lattice at $\beta=-22$ with 4 different initial configurations. Our best estimate of the critical $\beta$ for this volume is -22.09. 
The first initial configuration was completely ordered (in the $\beta\rightarrow -\infty$ sense, with $P$=1.5) by putting a nontrivial element of the center on a set of lines $\mathfrak{L}$. As we set $\beta = -22$, we expect to stay on the upper branch and end up with $P\simeq1.39$ (black dots in 
Fig. \ref{fig:meta}) for many iterations. 
The second configuration was completely random (empty circles) and stayed 
on the lower branch when $\beta$ was set to -22, to end up at $P\simeq1.34$. The third configuration (empty squares) was initially random, we then temporarily set $\beta =-27$ letting $P$ go up to 1.38, 
expecting to reach lower
metastable branch. When $\beta$ is set to -22, $P$ stabilized to the lower value 1.34. 
Finally, we prepared a fourth initial 
configuration (empty triangles) by first setting a nontrivial element of the center on a given $\mathfrak{L}$ and then temporarily setting $\beta=-17$ until $P$ is near 1.35 expecting to reach the upper 
metastable part. When $\beta$ is finally set to -22, we reach the upper branch 
value $P=1.39$. Fig. \ref{fig:meta} is quite similar to Fig. 1 of Ref. \cite{creutz81} and has the same type 
of crossings.

\begin{figure}
\includegraphics[width=3.4in,angle=0]{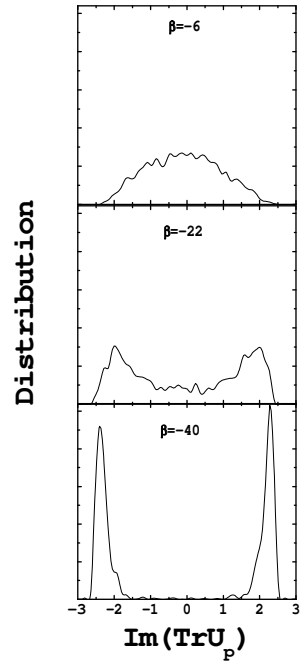}
\caption{Distribution of $ImTrU$ for three values of $\beta$.}
\label{fig:im}
\end{figure}

We believe that the first order transition observed above is similar to the one observed\cite{creutz79} for the $D=4$ $Z_2$ gauge 
theory. This model is dual to a nearest neighbor Ising model. In Fig. \ref{fig:im}, we show histograms of the 
distribution of $Im U$ below, near and above the transition. $ImU$ allows to separate the two nontrivial 
elements of the center $\Omega$ and $\Omega^{\dagger}$. As $\beta$ becomes more negative and goes through the
transition, a broad distribution around 0 develops two bumps which keep separating and sharpening as one would 
observe in an Ising model.

\begin{figure}[ht]
\includegraphics[width=2.8in,angle=0]{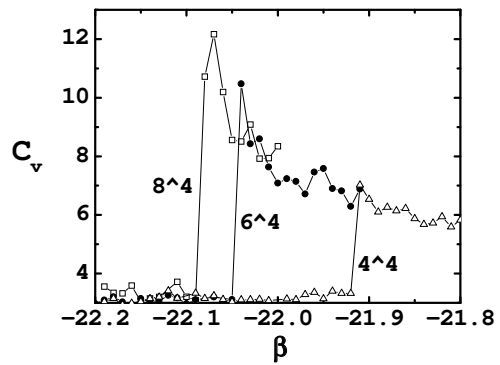}
\caption{The specific heat $C_V$ versus $\beta$ near the first order phase transition.}
\label{fig:cv}
\end{figure}

The transition can also be seen as a singularity in the specific heat defined in Eq. (\ref{eq:cv})
as shown in Fig. \ref{fig:cv}.
As expected the height of the peak increases with the volume. The location of the transition sightly 
moves left as the volume increases. 

\begin{figure}[ht]
\includegraphics[width=2.8in,angle=0]{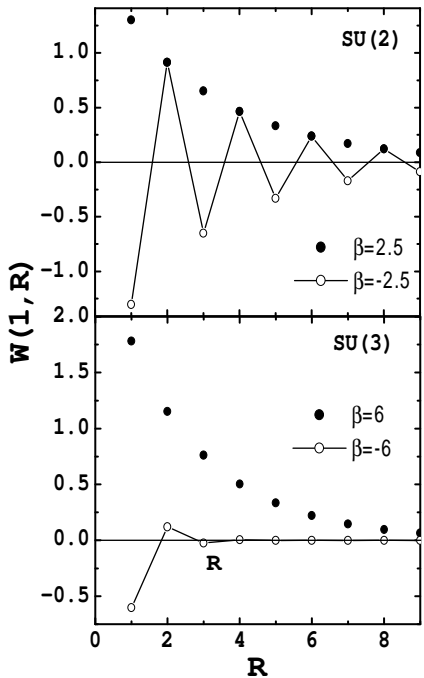}
\caption{The Wilson loop $\left\langle W(1,R)\right\rangle$ at $\beta = \pm 2.5 $ for $SU(2)$ and 
$\beta=\pm 6$ for $SU(3)$.}
\label{fig:2pic}
\end{figure}

Finally, we would like to compare the decay of the Wilson loop at negative $\beta$ for $SU(2)$ and $SU(3)$.
In Fig. \ref{fig:2pic}, we have plotted the Wilson loop $\left\langle W(1,R)\right\rangle$ for these two groups. 
For $SU(2)$ we observe the same decay as at positive $\beta$ but with alternated signs as predicted in 
section \ref{sec:su2}. For $SU(3)$, the decay is much faster than at the positive value of $\beta$ 
and show that the sign alternates as long as the signal is larger than the statistical fluctuations (namely 
for $R\leq 6$). 

The sign alternates at relatively low values of $|\beta|$. This agrees with the 
strong coupling expansion which predicts a $(-|\beta|/18)^{(R\times1)}$ behavior for 
$\beta$ negative and small in absolute value. 

\section{Conclusions}

We have studied lattice gauge theories at negative $\beta$. Wilson loops are well-defined 
and calculable with the Monte Carlo method. However, the limits $\beta \rightarrow \pm \infty$ 
of $P$ differ and an expansion in $g^2\propto 1/\beta$ cannot have a finite radius of convergence. 
This statement has been substantiated  for $N=2$ and 3, but from the discussion of 
section \ref{sec:limit}, it seems clear that it should extend to general $N$.

We found a first order phase transition near $\beta = -22$.
At this point, it
seems unrelated to the known transition near 
$\beta \simeq +6$ and branch cuts in the complex plane discussed by J. Kogut \cite{kogut80}. However, a 
more complete picture may appear if we study $P$ 
for a larger class of action. It is conceivable that by introducing a 
linear combination of terms involving larger contours than the $1\times 1$ plaquette, multiplied by one free 
parameter, we could create a line of first order phase 
transition ending at a second order phase transition. 
Another possibility would be to add an adjoint term as in Ref. \cite{bhanot81}. 
Finding new second order phase transitions would allow us to define a nontrivial continuum limit. 
This could be of interest in the 
context of cosmology and stellar evolution.

\begin{acknowledgments}
This research was supported in part by the Department of Energy
under Contract No. FG02-91ER40664. We thanks C. Bender, M. Creutz, M. Ogilvie and the participants of the Argonne workshop ``QCD in extreme environment'' for valuable conversations.
\end{acknowledgments}
\appendix*
\section{
Maximal sets of Non-intersecting of lines on a cubic lattice}

In this Appendix, we  consider a $D$ dimensional cubic lattice. 
We restrict the use of ``line'' to collections of links along the $D$ principal directions of the lattice and the use of ``plane'' to collections of plaquettes along the $D(D-1)/2$ principal orientations. 
In other words, these objects are lines and planes in the usual sense, but we 
exclude some ``oblique'' sets that can be constructed out of the sites.

We now try to construct a set of lines such that every plaquette shares one and only one link with this set. It is obvious that these lines cannot intersect, otherwise, 
at the point of intersection and in the plane defined by the two lines, we could fit 
4 plaquettes, each sharing two links with the lines. These lines cannot be obtained from 
each other by a translation of one lattice spacing in one single direction, otherwise 
the set of lines would share two opposite links on the plaquettes in between the two lines.
  
For $D=2$, the problem has obvious solutions, we can pick for instance a set of vertical lines separated by two lattice spacings. 
Using translation by one lattice spacing and rotation by $\pi/2$, it is possible to obtain three other solutions.
For $D>2$, it is sufficient to show 
that for every plane (in the restricted sense defined above), we have a $D=2$ solution. As this restricted set of planes contains all the plaquette ounce, we would have then succeeded in proving the assertion. If such a solution exists, it is invariant by a translation by 2 lattice spacing in any direction. Consequently, we only need to prove the existence of the set of lines on a $2^D$ lattice with periodic boundary conditions. The full solution is then obtained by translation of the $2^D$ solution. If the lattice is finite, this only works if $L$ is even, an assumption 
we have maintained in this article.

On a $2^D$ lattice, the lines (as defined above) are constructed by fixing $D-1$ 
coordinates values to be 0 or 1 and leaving the remaining coordinate arbitrary.
For instance, for $D=3$, a line in the 3rd direction coming out of the origin 
will be denoted $(0,0,A)$ where $A$ stands for arbitrary and means 
0 or 1. In general $D$, there are $D2^{D-1}$ such lines. Consequently, there are $D2^{D}$ links, each shared 
by $2(D-1)$ plaquettes. There are thus $D2^D2(D-1)/4=D(D-1)2^{D-1}$ plaquettes. A set of lines which 
has exactly one link in common with every plaquette, has $D(D-1)2^{D-1}/(2(D-1))=D2^{D-2}$ links in other words 
it must contain $D2^{D-3}$ lines. For $D=2$, such a set has only one line and there are four possible choices.
For $D=3$, an example of solution is 
$\{(A,0,0),(0,A,1),(1,1,A)\}$. It is not difficult to show that there are 8 distinct solutions of this type.
For $D=4$, a solution consists in 8 lines. An example of solution is 
\begin{eqnarray}
\nonumber
&\{&(A,0,0,0),(0,A,0,1),(0,1,A,0),
(0,0,1,A),\\ \nonumber &\ &(1,1,0,A)
 ,(1,0,A,1),(1,A,1,0),(A,1,1,1)\ \}\ .
\end{eqnarray}


\end{document}